\documentclass[usenatbib]{mn2e}

\usepackage[dvips]{epsfig}

\bibpunct{(}{)}{;}{a}{}{,}
\def\dt{\mathrm}

\begin{document}

\title[GRB Host Galaxies]{Host Galaxies of Gamma-Ray Bursts 
  and their Cosmological Evolution}

\author[Courty, Bj\"ornsson \& Gudmundsso]
    {St\'ephanie Courty, Gunnlaugur Bj\"ornsson, Einar H.\ Gudmundsson\\
    Science Institute, University of Iceland, Dunhaga~3, 
    IS--107 Reykjavik, Iceland\\ {\tt e-mail: courty, gulli, einar@raunvis.hi.is}}

\date{\today}


\maketitle

\begin{abstract}
We use numerical simulations of large scale structure formation to
explore the cosmological properties of Gamma-Ray Burst (GRB) host
galaxies. Among the different sub-populations found in the
simulations, we identify the host galaxies as the most efficient
star-forming objects, i.e. galaxies with high specific star formation
rates. We find that the host candidates are low-mass, young galaxies
with low to moderate star formation rate. These properties are
consistent with those observed in GRB hosts, most of which are
sub-luminous, blue galaxies. Assuming that host candidates are
galaxies with high star formation rates would have given conclusions
inconsistent with the observations. The specific star formation rate,
given a galaxy mass, is shown to increase as the redshift
increases. The low mass of the putative hosts makes them difficult to
detect with present day telescopes and the probability density
function of the specific star formation rate is predicted to change
depending on whether or not these galaxies are observed.
\end{abstract}



\section{Introduction}

For three decades Gamma-Ray Bursts (GRBs) remained an enigma. The
detection of optical afterglows associated with a GRB \citep[see][for
a review]{vanPar2000}, revolutionized our understanding of the
phenomena. A GRB is now thought to occur when the core of a massive
star collapses to form a black hole \citep{Woosley1993, Pac1998,
  MacFadyen1999} \citep[see also][for a review]{Mesz2002}.  In most
cases when an optical afterglow from a GRB has been detected, they
have been shown to occur in galaxies at intermediate or high
redshifts.  To date, the redshift of over 40 GRB afterglows and their
host galaxies have been measured.  Because of the short lifetime of a
massive star, a GRB is generated essentially instantaneously after the
star's formation, as compared to the evolutionary timescale of a
typical galaxy or the cosmological timescale.  The bursts are
therefore generally considered to be good tracers of the history of
massive star formation \citep[e.g.][]{Blain2000}.

It was early realized that because of their extreme brightness, GRBs
and their afterglows might be detectable to very high redshifts and
therefore might be useful as cosmological probes, in particular if the
bursts possessed a 'standard candle' like property
\citep[e.g.][]{Lamb2000}.  The main interest in the cosmological
application of GRBs has resulted from the increased understanding of
the relation between GRBs and powerful supernovae
\citep[e.g.][]{Hjorth2003, Stanek2003} and from there the mapping of
the cosmic star formation history \citep[e.g.][and references
therein]{Ram-Ruz2002}.  Other possible applications have also been
considered, such as the use of bursts as probes of the interstellar
medium in evolving galaxies and as probes of primordial star formation
and reionization, although these are observationally very challenging
at present \citep[e.g.][]{Djorg2003}.  Constructing a reliable Hubble
diagram with GRB afterglows, however, is likely to remain a very
difficult subject \citep{Bloom2003}.

In this paper, we take a different approach altogether. Instead of
focusing on the diverse properties of the bursts themselves we use the
fact that in almost all cases where sufficiently deep observations
have been carried out, a host galaxy has been discovered. Many of the
host galaxies are very faint, with $R$ band magnitudes down to about
30 \citep{Jaunsen2003}, and might have gone undetected had not a GRB
occurred in them. We will occasionally refer to the hosts as {\em GRB
  selected} galaxies.  Unfortunately, the number of hosts is still too
limited for statistical studies, but this is likely to change after
{\em Swift} is launched in 2004. A number of first results on the
hosts does exist however, and a picture of their properties is
beginning to emerge. The host morphology shows most of them to be
dwarf galaxies, either compact, irregular or interacting, possibly
merging, and they also tend to be blue in color
\citep[e.g.][]{Bloom1998, LeFloch2003, Berger2003}. Many hosts are
sub-luminous, with $L/L_*<1$, \citep[e.g.][]{LeFloch2003}, with
GRB~990705 a notable exception \citep{LeFloch2002}. The star formation
rate ($SFR$) in most hosts as inferred from optical observations show
low to intermediate activity, with $SFR=1-50 M_{\sun}$/yr
\citep[e.g.][]{Bloom1999, Castro-Tirado2001, Fynbo2001, Chary2002,
  Price2002}.  In a number of cases the sub-mm and radio observations
indicate a $SFR$ that is up to a factor of 10 higher than the
optically inferred rates \citep{Berger2003}.  Finally,
\cite{Sokolov2001} and \cite{Chary2002} infer a host galaxy mass, $M$,
typically in the range $2\cdot 10^{8}-4\cdot 10^{10}M_{\sun}$.  In
addition, the variation in the specific star formation rate, $SFR/M$,
from host to host is modest, normally within a factor of a few
\citep{Chary2002, Christensen2004}. In fact, \cite{Christensen2004}
has shown that all the hosts in their sample have spectral energy
distributions similar to young starburst galaxies with small to
moderate extinction. Comparing their sample with the Hubble Deep
Field, they find the hosts to be similar to the bluest of the field
galaxies which have the highest specific $SFR$.

As the bursts occur in galaxies, they can be viewed not only as
tracers of massive star formation, but also as tracers of galaxy
formation and evolution.  The study of GRBs and their host galaxies is
therefore intimately linked to the study of galaxy evolution in
general. Although the presently known host sample indicates that
bursts can occur in any type of galaxy, the majority of them are faint
and show a dwarf-like morphology. They may therefore belong to the
faint end of the galaxy luminosity function. GRB selected galaxies may
prove to be an important sub-population of the galaxy population and
their study may thus provide information about the galaxy luminosity
function not easily accessible otherwise.

The aim of this paper is to investigate theoretically the properties
of the host galaxies of GRBs as well as their cosmological evolution
rather than to focus on the GRB progenitor and its properties. To this
end, we use numerical simulations of large scale structure formation
with the aim of identifying objects in the simulations with properties
corresponding to those observed in the hosts. The simulations we use,
were previously considered by \cite{AC2004} to explore a number of
galaxy properties such as the cosmic star formation rate density and
the galaxy mass function. The simulations reveal a significant
population of low-mass galaxies whose existence raises a number of
questions on galaxy evolution. In this paper we concentrate on the
instantaneous and the specific star formation rates of the simulated
galaxies.  We identify a sub-population of the galaxies with high
specific star formation rate as the likely candidates for GRB hosts.
We find that the majority of this putative host population has
low-mass and low to modest $SFR$, although a number of high-mass
objects also have similar specific rates.  We then explore the
properties and cosmic evolution of the simulated host population and
compare to the galaxy population as a whole.

In section~\ref{def}, we describe the details of the numerical
simulations relevant for this study. In section~\ref{lowz} we
introduce the instantaneous star formation rate and in particular the
specific star formation rate as a measure of the efficiency of star
formation. We also discuss the properties of the general galaxy
population at low redshift.  In section~\ref{evol} a similar
discussion is presented for high redshift and the cosmological
evolution of the galaxy properties is explored.  In section
\ref{discuss} we propose interpretations of the observations and we
argue that GRB host galaxies may be a powerful tracer of the faint
galaxy population. Section \ref{conclude} concludes the paper.

\section{Numerical simulations}
\label{def}

The simulations were performed with a 3D N-body/hydro\-dynamical code,
coupling a PM scheme for computing gravitational forces with an
Eulerian approach for solving the hydrodynamical equations. Shock
heating is treated with the artificial viscosity method \citep{VNR}.
The radiative cooling processes included are: collisional excitation,
collisional ionization, recombination, bremsstrahlung and Compton
scattering. Collisional ionization equilibrium is not assumed and the
cooling rates are computed from the evolution of a primordial
composition hydrogen-helium plasma. The simulations also include a
model for galaxy formation, summarized below. The numerical procedures
are described in more detail in \cite{CA2004} and \cite{AC2004}.  The
galaxy properties discussed in the latter paper were found to be
consistent with observational data. This suggests that the galaxy
formation model used here to investigate the nature of host galaxies
of GRBs, captures the essence of the galaxy formation process.

Numerical simulations of large scale structure formation do not
currently allow following the formation of objects of mass lower than
the scale of a grid cell, a few times $10^6 \ M_{\odot}$. But
computing the thermodynamic properties of the gas and its cosmological
evolution allow us to consider the physical conditions needed to form
a galaxy. The most important condition is that the gas cloud is
collapsing, in the sense that the cooling time is less than the
dynamical time or the free fall time \citep{Rees77}. We then identify,
at each time step, gas regions satisfying this condition. A fraction
of the gas mass is turned into an object, labeled a ``stellar
particle'', with a rate:
\begin{equation}
\frac{\textrm{d}m_B}{\textrm{d}t}=-\frac{m_B}{t_*},
\label{eq1}
\end{equation}
with $m_B(t)$ the available baryonic mass enclosed within the gas
region at cosmic time $t$, and $t_*$ a characteristic timescale. The
cosmological evolution of these particles is followed in a
non-collisional way. To express the condition $t_{\rm cool} < t_{\rm
  ff}$, we define the cooling timescale $t_{\rm cool}$, computed from
the internal energy variation of the gas, and the dynamical time or
free fall time, $t_{\rm ff} = \sqrt{3 \pi/32 G \rho}$. To make sure
that gas regions giving birth to galaxies are correctly identified we
use additional criteria: the size of the gas cloud must be less than
the Jean's length given by $\lambda_J = c_s (\pi/G \rho)^{1/2}$; the
gas must be in a converging flow: $\nabla \cdot \vec v < 0$; the
baryonic density contrast, $\delta_B \equiv (\delta \rho
/\bar{\rho})_B$, must be higher than a threshold $(1+\delta_B)_s$,
here taken to be the value of the baryonic density contrast at the
turnaround, i.e.\ 5.5 as computed in the top-hat collapse spherical
model \citep{Padmanabhan}. The total matter density, including dark
matter, baryonic matter and stellar particles, is used in the
expressions for the dynamical time and the Jean's length.  In each
cell, checking that the four criteria described above are satisfied, a
fraction $\Delta t/t_*$ of the gas is turned into a stellar particle,
with $\Delta t$ the timestep of the simulation.  If a stellar particle
starts forming at cosmic time $t_0$ when the baryonic mass enclosed
within the grid cell is $m_B(t_0)$, we assume that the particle is
fully formed at time $t_i=t_0+\Delta t$. Using eq.~(\ref{eq1}), we
find that the final mass of the particle is given by
\begin{equation}
\label{mstar}
m_i = m_B(t_0) - m_B(t_i) \simeq m_B(t_0) \frac{\Delta t}{t_*}.
\end{equation}
The characteristic time is taken to be
$t_*=\dt{max}(t_{\rm ff},10^8 \ \dt{yr})$. The stellar particles are
involved in the computation of the gravitational potential and their
evolution is treated in the same way as the collisionless dark matter.

Galaxy-like objects are then defined, at any redshift, by grouping the
stellar particles with a friend-of-friend algorithm. This algorithm
joins together all particles closer than a distance proportional to a
link parameter $\eta$, usually taken to be 0.2 (equivalent to
selecting an overdensity of 187 in the spherical collapse model). The
mass range of the stellar particles spans from $10^6$ to $10^8$
$M_{\sun}$, but in what follows we only consider, at any redshift,
galaxies with a total mass higher than $5\cdot 10^8 M_{\sun}$,
although a number of lower mass galaxies also form in the simulations.
At $z=0$, this lower galaxy mass limit corresponds to a minimum of 11
stellar particles per galaxy and at $z=3$ it corresponds to a minimum
of 27 particles. The total number of galaxy objects in the catalogs at
$z=0$, 1 and 3 are 1650, 2164 and 1602, respectively.

\begin{figure}
\begin{center}
  \includegraphics[height=8.5cm, angle=-90]{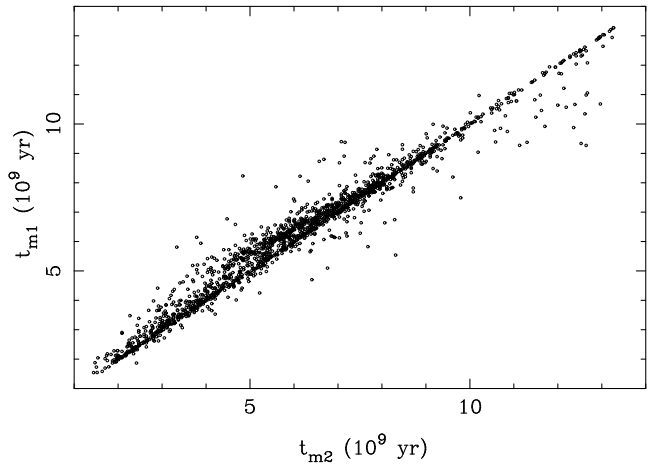}
\caption{Comparing the two different ways of estimating the epoch of 
  formation of galaxies, $t_{\rm m1}$ and $t_{\rm m2}$, computed from 
  the galaxy catalog at $z=0$. See text for definitions.}
\label{fig1}
\end{center}
\end{figure}

Through the formation of the stellar particles and their groupings we
therefore have access to the formation history of each galaxy at any
redshift. The mass of the galaxy at any time is the sum of the mass of
its individual stellar particles at that time. This mass is likely to
be an estimate of all the stellar populations formed over the lifetime
of the galaxy, as the dark matter component is not included in the
galaxy mass. We estimate the epoch of formation for a given galaxy as
the mean epoch of formation of all stellar particles the galaxy is
composed of, weighted by the mass of each particle, i.e.\ $t_{\rm
  m1}=\sum (m_i t_i)/\sum m_i$, where $t_i$ is the cosmic time
corresponding to the epoch of formation of stellar particle $i$. An
alternative method would be to define the formation epoch as the
epoch, $t_{\rm m2}$, at which the galaxy has acquired half of its
final mass.  Both methods give similar epoch estimates as shown in
Figure~\ref{fig1}. As the former method is based on the formation
history of the stellar populations that make up the galaxy, it
resembles the way galaxy ages are estimated observationally, namely
from the age of the stellar populations in the galaxy. In this work,
all formation epochs are therefore estimated with the first method.

The results of this paper are given for a $\Lambda-$cold dark matter
model. The parameters of the simulations are: $H_0=70\ \dt{km} \ 
\dt{s}^{-1} \ \dt{Mpc}^{-1}$, $\Omega_K=0$, $\Omega_m=0.3$,
$\Omega_{\Lambda}=0.7$, $\Omega_b=0.02h^{-2}$ with $h=H_0/100$. The
initial density fluctuation spectrum uses the transfer functions taken
from \cite{Bardeen} with a shape parameter given by \cite{Sugiyama}.
The fluctuation spectrum is normalized on COBE data \citep{Bunn}
leading to a filtered dispersion at $R=8 \ h^{-1} \ \textrm{Mpc}$ of
$\sigma_8=0.91$. The resolution of the simulations is as follows: the
number of dark matter particles is $N_p=256^3$ and the number of grid
cells is $N_g=256^3$; the comoving size of the computational volume is
$L_{box}=32 \ h^{-1}Mpc$, giving a dark matter particle mass of
$2.01\times 10^8 \ M_{\odot}$, and a gas mass initially enclosed in
each grid cell of $3.09\times 10^7 \ M_{\odot}$.

\section{Instantaneous star formation rate and efficiency of star formation}
\label{lowz}

\begin{figure}
\begin{center}
\includegraphics[height=8.cm, angle=-90]{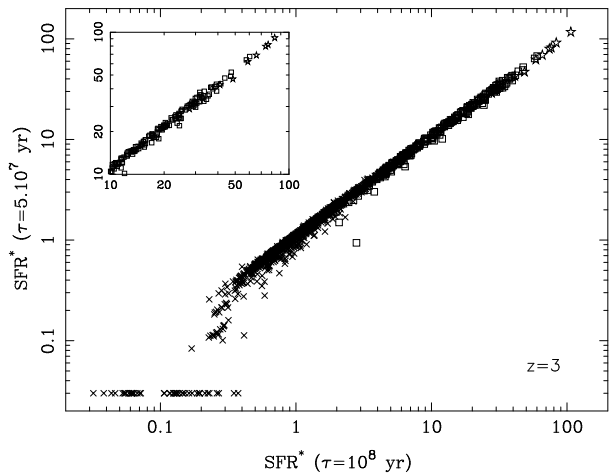}
\caption{Comparison between the instantaneous star formation rates, in 
  units of $M_{\sun}$/yr, computed for time dilated values of
  $\tau=0.1$ Gyrs and $\tau=0.05$ Gyrs for the catalog at $z=3$.
  Galaxies with $SFR^{*}=0.0$ at $\tau=0.05$ Gyrs are plotted at
  $SFR^{*}=0.03$. Symbols denote the mass range of each galaxy:
  $5\cdot 10^{8} - 5\cdot 10^{9}M_{\sun}$ (crosses), $5\cdot 10^{9} -
  10^{11}M_{\sun}$ (squares), $M>10^{11}M_{\sun}$ (stars).  The inset
  shows a blow-up of the highest $SFR^{*}$-range. }
\label{fig2}
\end{center}
\end{figure}

For each object at any given redshift $z$, corresponding to cosmic
time $t_z$, we compute the instantaneous star formation rate $SFR^*$,
where $*$ indicates a value obtained from the simulations.  The
instantaneous rate is computed from the mass in stellar populations in
the galaxy that have age less than $\tau$ at time $t_z$, $SFR^*=\Delta
M/\tau$. This is equivalent to the baryonic mass $\Delta M$, having
been converted to stars during time $\tau$ prior to the epoch under
consideration.  Note that the $SFR^*$ depends both on $\tau$ and the
cosmic time, $SFR^*(\tau,t_z)$. It depends also on the amount of gas
in the galaxy that is available for star formation at $t_z$ since the
star formation rate in the model is proportional to the baryonic
density. We use $\tau=0.1$ Gyrs as representative value at $z=0$, time
dilated at higher redshifts. Figure~\ref{fig2} shows a comparison
between $SFR^*$ computed with $\tau=0.05$ and $\tau=0.1$ Gyrs for
galaxies in the catalog at $z=3$. Also indicated in the figure are the
masses of the objects as explained in the caption. We note that the
star formation rates appear to be rather insensitive to the adopted
value of $\tau$, indicating that for most of the simulated galaxies,
the $SFR^*$ is rather constant over a timescale of 0.1 Gyrs at high
redshift (see also \cite{Weinberg2002} for similar comparisons). There
is some dispersion in the two estimates, mainly due to irregularities
in the star formation events in a given galaxy. In addition, some
galaxies that have a non-zero $SFR^*$ as computed with $\tau=0.1$
Gyrs, show no star formation on shorter timescales. These galaxies are
also shown in fig.~\ref{fig2}, but assigned an arbitrary value of
$SFR^*(\tau=0.05)=0.03 \ M_{\sun}$/yr.  For the low-mass objects with
the lowest $SFR^*$, there is an apparent trend for the lower value of
$\tau$ to yield a lower value of $SFR^*$, up to a a factor of a few.
The larger value of $\tau$ samples a longer period of star formation
in the galaxy prior to the epoch under consideration and may therefore
include periods of stronger star formation activity, resulting in a
higher $SFR^*$.

\begin{figure}
\begin{center}
\includegraphics[height=8.3cm, angle=-90]{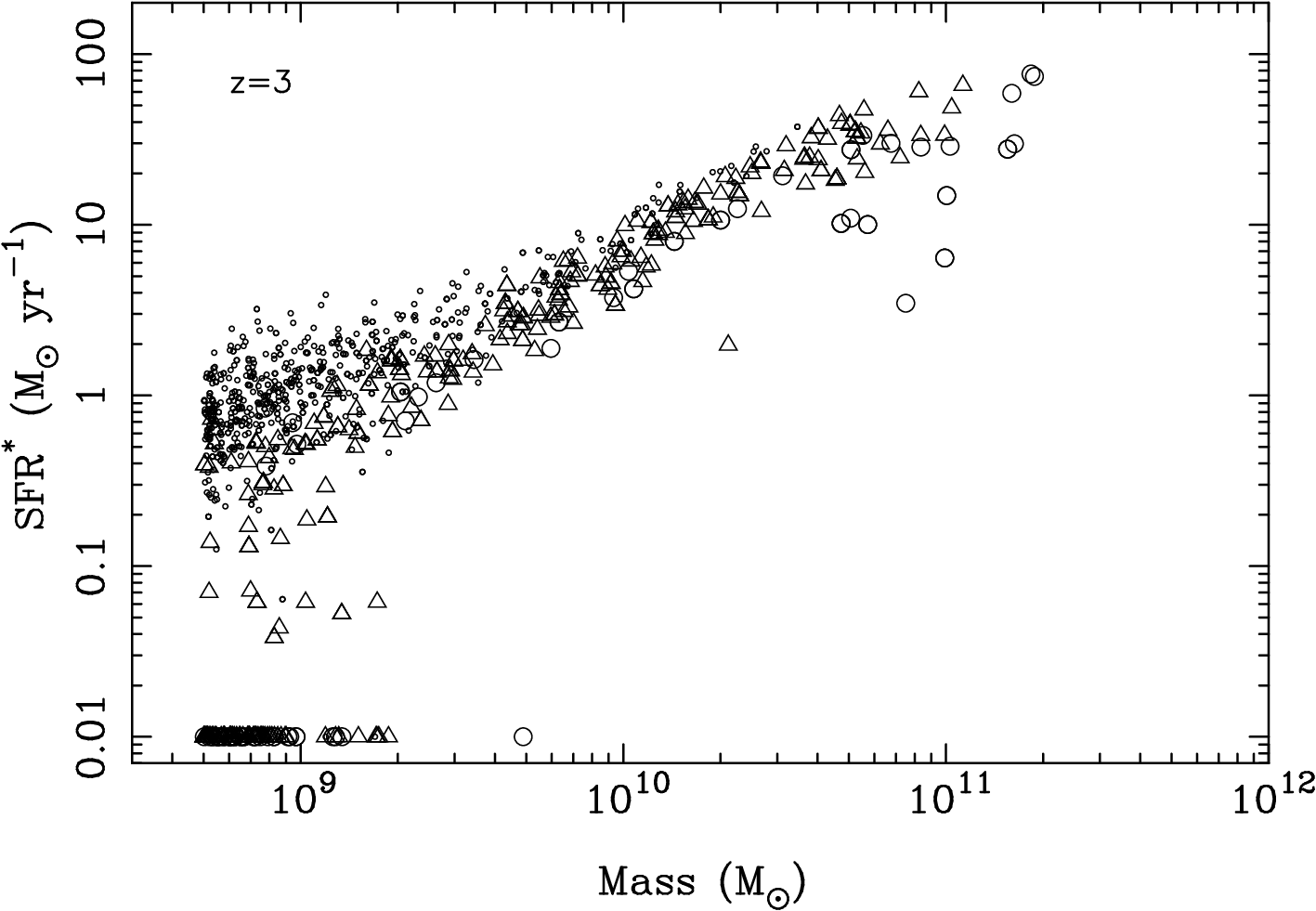}
\includegraphics[height=8.3cm, angle=-90]{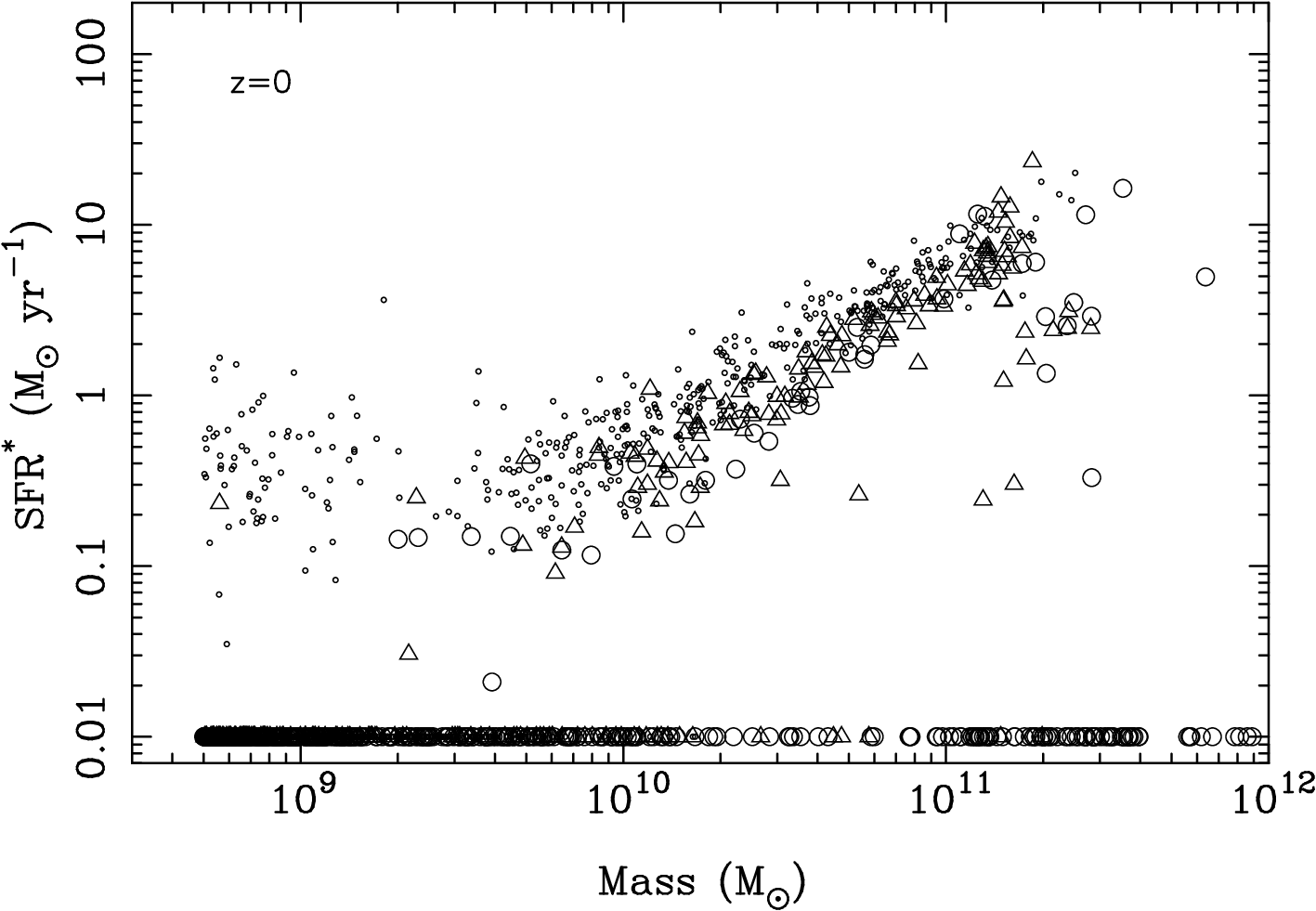}
\caption{ Instantaneous star formation rate as a function of galaxy mass,
  computed with $\tau=0.1$ Gyrs, for the catalogs at $z=3$ (upper
  panel) and $z=0$ (lower panel). For clarity, only 1000 objects with
  non-zero $SFR^*$ are plotted from the catalog at $z=3$. The symbols
  indicate the epoch of formation of each galaxy. In the upper panel
  they denote: epoch of formation less than $z=3.9$ (dots), epochs in
  the range $z=3.9-4.3$ (triangles) and epochs greater than $4.3$
  (open circles). In the lower panel we have: less than $z=0.9$
  (dots), in the range $z=0.9-1.1$ (triangles) and greater than
  $z=1.1$ (open circles). In both panels galaxies with $SFR^{*}=0.0$
  are plotted at $SFR^{*}=0.01 \ M_{\sun}$/yr.}
\label{fig3}
\end{center}
\end{figure}

Figure~\ref{fig3} shows the $SFR^*$ as a function of the mass of the
objects for the catalogs at $z=0$ and $z=3$. Galaxies with non-zero
$SFR^*$ are plotted arbitrarily at $SFR^{*}=0.01 \ M_{\sun}$/yr. In
the former catalog, 572 objects have non-zero $SFR^*$ for $\tau=0.1$
Gyrs, while in the latter these objects are 1470, but for clarity only
1000 randomly selected objects with non-zero $SFR^*$ are plotted. Also
indicated in fig.~\ref{fig3} are three ranges of the epoch of
formation of the galaxies. For the catalog at $z=0$, the dividing
epochs are $z=0.9$, and $z=1.1$, roughly corresponding to half and the
third of the age of the Universe. For the catalog at $z=3$, we use
$z=3.9$ and $z=4.3$ as the dividing epochs. For both redshifts, the
instantaneous star formation rate increases linearly with mass
although there is considerable dispersion especially at the lower
redshift. For $z=0$, this linearity holds only in the approximate mass
range $10^{10}-10^{11}M_{\sun}$, for lower masses there is a large
scatter in the $SFR^*$. For the higher redshift the linearity extends
over two decades in mass.  It is clear at both redshifts that the
older objects tend to have lower $SFR^*$ than the younger objects and
a slightly steeper mass dependence than linear, though the difference
decreases with increasing redshift.

Note also, that galaxies of a given mass tend to have up to 10 times
higher $SFR^*$ at $z=3$ than at $z=0$ and that the maximum $SFR^*$ in
the catalogs decreases with redshift from about 100 $M_{\sun}$/yr at
$z=3$ to about 10-20 $M_{\sun}$/yr at $z=0$. It is interesting to note
that these high $SFR^*$ values at high redshift are consistent with
the observationally inferred $SFR$ of Lyman-break galaxies at $z=3$
\citep{Giaval2002}. The evolution of $SFR^*$ with redshift is very
weak for the low-mass objects in fig.~\ref{fig3}, as their $SFR^*$
tends to stay constant in time. The most dramatic difference in the
cosmological evolution of the $SFR^*$ is seen in high-mass, old
galaxies. The oldest (open circles) and most massive objects have a
$SFR^*$ up to two orders of magnitudes lower than their younger
counterparts (dots). In fact, the star formation rate of massive,
early-formed galaxies at $z=0$ is similar to that of the population of
low-mass objects (few times $10^9M_{\sun}$ or less), almost all of
which are late-formed galaxies. 

Note that the plots in fig.~\ref{fig3} show a number of galaxies which
are not forming stars at the time of observation, $t_z$. In these
objects star formation has already ceased or may be temporally
stopped.  These objects, although inactive, are quite numerous with
most of them being of low-mass at both redshifts but being of higher
mass as the redshift decreases. At low redshift, most of the massive
galaxies show no activity at all.  These high-mass, inactive galaxies
reside in the center of the most massive dark matter halos
\citep{AC2004} and, as the evolution progresses, the depth of the
potential wells prevents more and more gas from cooling.  This
hierarchical character of the galaxy formation process results in a
strongly reduced star formation activity at low redshift, and
resembles that observed in elliptical galaxies in the center of galaxy
clusters. Adopting a higher value of $\tau$ would indeed cause some of
these massive, early-formed galaxies to have a non-zero $SFR^*$
(recall fig.~\ref{fig2}) as it would take into account the latest
formed stellar populations, in the case where there is still some cold
gas available. This inactive population is important for our
understanding of galaxy evolution and will be discussed further in
Section \ref{discuss}.

\begin{figure}
\begin{center}
\includegraphics[height=8.3cm, angle=-90]{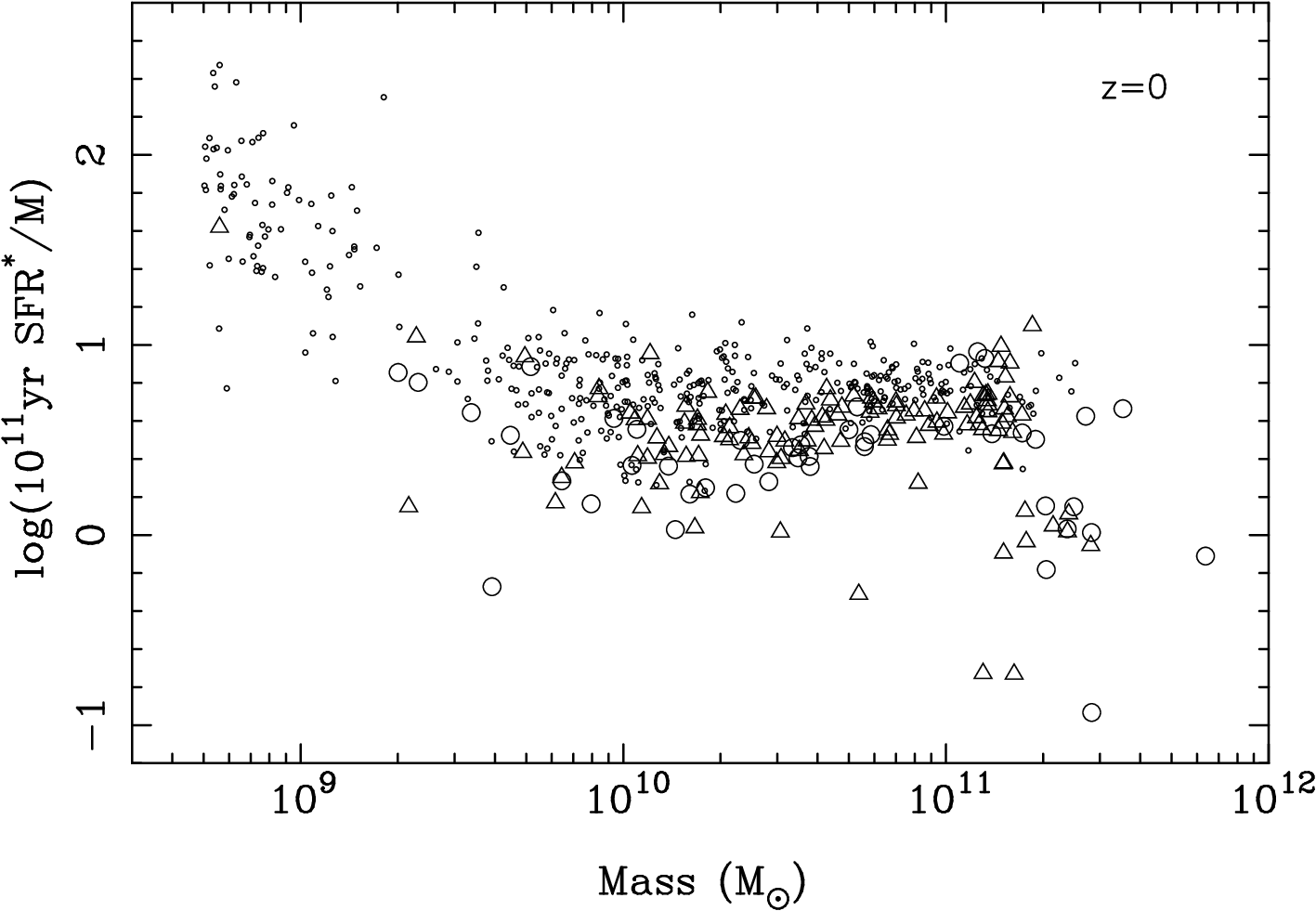}
\includegraphics[height=8.cm, angle=-90]{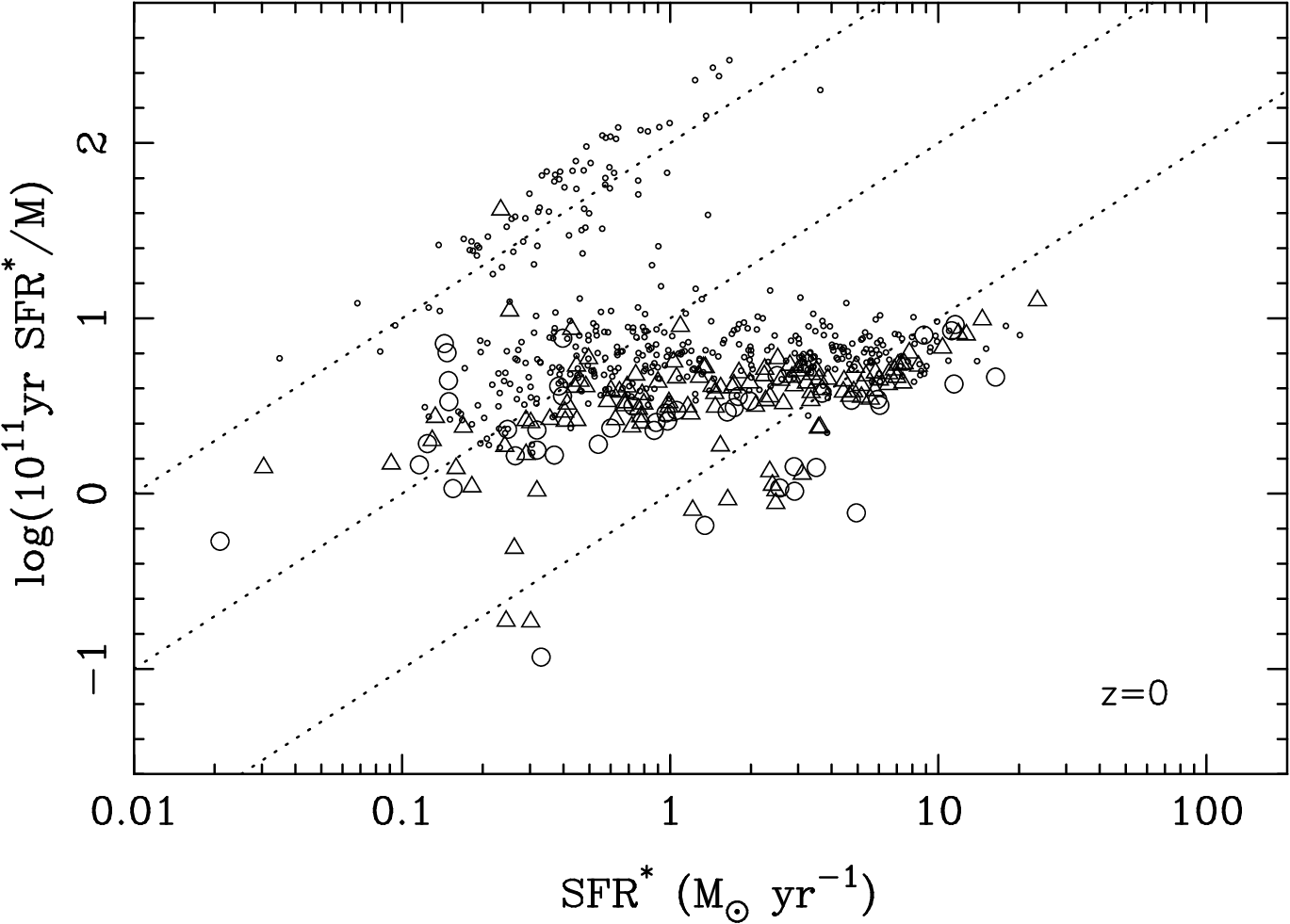}
\caption{Upper panel: Efficiency parameter 
  $\epsilon$, as a function of galaxy mass for the catalog at $z=0$.
  Compare also with fig.~\ref{fig3}, where the different symbols are
  explained.  Lower panel: $\epsilon$ as a function of $SFR^*$ for
  galaxies in the catalog at $z=0$.  The dotted lines indicate a
  constant mass of $10^9, 10^{10}$ and $10^{11}M_{\sun}$ (from left to
  right).}
\label{fig4}
\end{center}
\end{figure}

The amount of gas turned into stars at any given epoch, $t_z$, depends
on the amount of cold gas available in a gas region, hence the scaling
of the $SFR^*$ with the mass of a galaxy. However, a low-mass galaxy
can be more efficient in turning gas into stars than a high-mass
galaxy. A measure of this efficiency is given by the specific star
formation rate, the ratio of the $SFR^*$ to the mass of the galaxy.
The specific star formation rate is usually normalized to a given mass, 
and we define a star formation efficiency parameter $\epsilon$ by,
\begin{eqnarray}
\epsilon & \equiv & \log\left(\frac{SFR^*}{M_{\sun} {\rm yr^{-1}}}\right)
               \left(\frac{M}{10^{11}M_{\sun}}\right)^{-1}\nonumber \\
         & = & \log(10^{11} {\rm yr}\; SFR^*/M).
\end{eqnarray}
Although $\epsilon$ is a dimensionless parameter we will use it and
the specific star formation rate, interchangeably and with the same
meaning.  In fig.~\ref{fig4}, we plot $\epsilon$ as a function of mass
in the upper panel and as a function of $SFR^*$ in the lower panel,
for galaxies in our catalog at $z=0$. In the lower panel, galaxies of
a given mass are located along diagonal lines, three of which are
shown.  It is apparent that the low mass galaxies have a much higher
specific star formation rate than intermediate or high-mass galaxies
and are therefore much more efficient in turning baryonic mass into
stellar mass. In fact, the old and very high-mass objects, $M>10^{11}
\ M_{\sun}$, and some of the older intermediate mass objects, have
specific rates that are up to two orders of magnitudes lower than the
low-mass objects ($M< 10^{10} \ M_{\sun}$).  The results presented in
the upper panel show a similar trend as seen in observational studies
on blue compact galaxies \citep{Guzman1997} and star-forming galaxies
\citep{PerGon2003}, where the specific star formation rate is seen to
increase with decreasing galaxy mass for low-mass galaxies, $M<
10^{10} \ M_{\sun}$. Despite approximations and a number of
uncertainties inherent in numerical simulations, we find that our
specific rates agree quite well with the observationally inferred
values. The low-mass objects are observed to have $\epsilon$ in the
range $1.25-3$ \citep{Guzman1997}, where we find the range to be about
$1-2.5$ in the simulations.

The lower panel of fig.~\ref{fig4} reveals several interesting
properties of the galaxy population.  It is apparent that the galaxies
tend to cluster in different regions in this figure. Most of the
objects in the intermediate mass range, have an efficiency in the
range $\epsilon=0-1$, although there is a clear difference between the
young and the old objects, the younger ones in general being more
efficient.  The star formation activity of these intermediate mass
objects, as defined by their star formation rate, spans the range
$SFR^*=0.1-20 \ M_{\sun}$/yr.  Another interesting feature seen in
fig.~\ref{fig4} are the two distinct groups of objects, one of
high-mass with low efficiency ($\epsilon\la 0$), and the other of
low-mass but with high efficiency ($\epsilon>1$). The low-mass
objects, that cluster around the diagonal line representing a galaxy
mass of $10^9M_{\sun}$, have the highest specific star formation rate,
but only an $SFR^*$ in the approximate range $0.1-1 \ M_{\sun}$/yr.
The most massive objects, although again spanning almost two orders of
magnitude in their $SFR^*$, distinguish themselves by having by far
the lowest specific star formation rate.  In general terms, we thus
see that the low-mass objects are the most efficient in their star
formation, by a factor of 10 over the intermediate class. The most
massive ones are again a factor of at least $5-10$ less efficient than
the intermediate ones.

\begin{figure}
\begin{center}
\includegraphics[height=8.3cm, angle=-90]{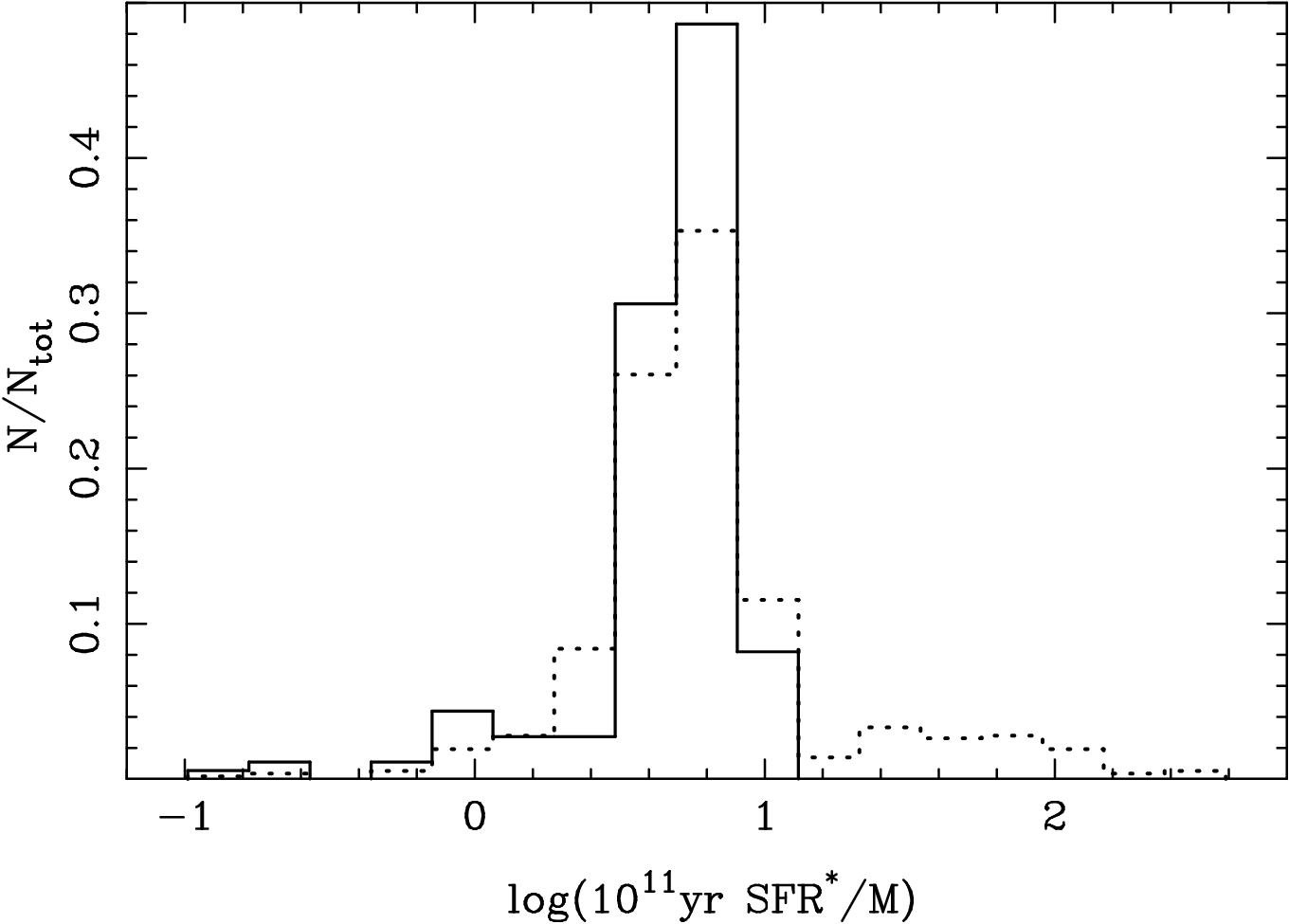}
\includegraphics[height=8.3cm, angle=-90]{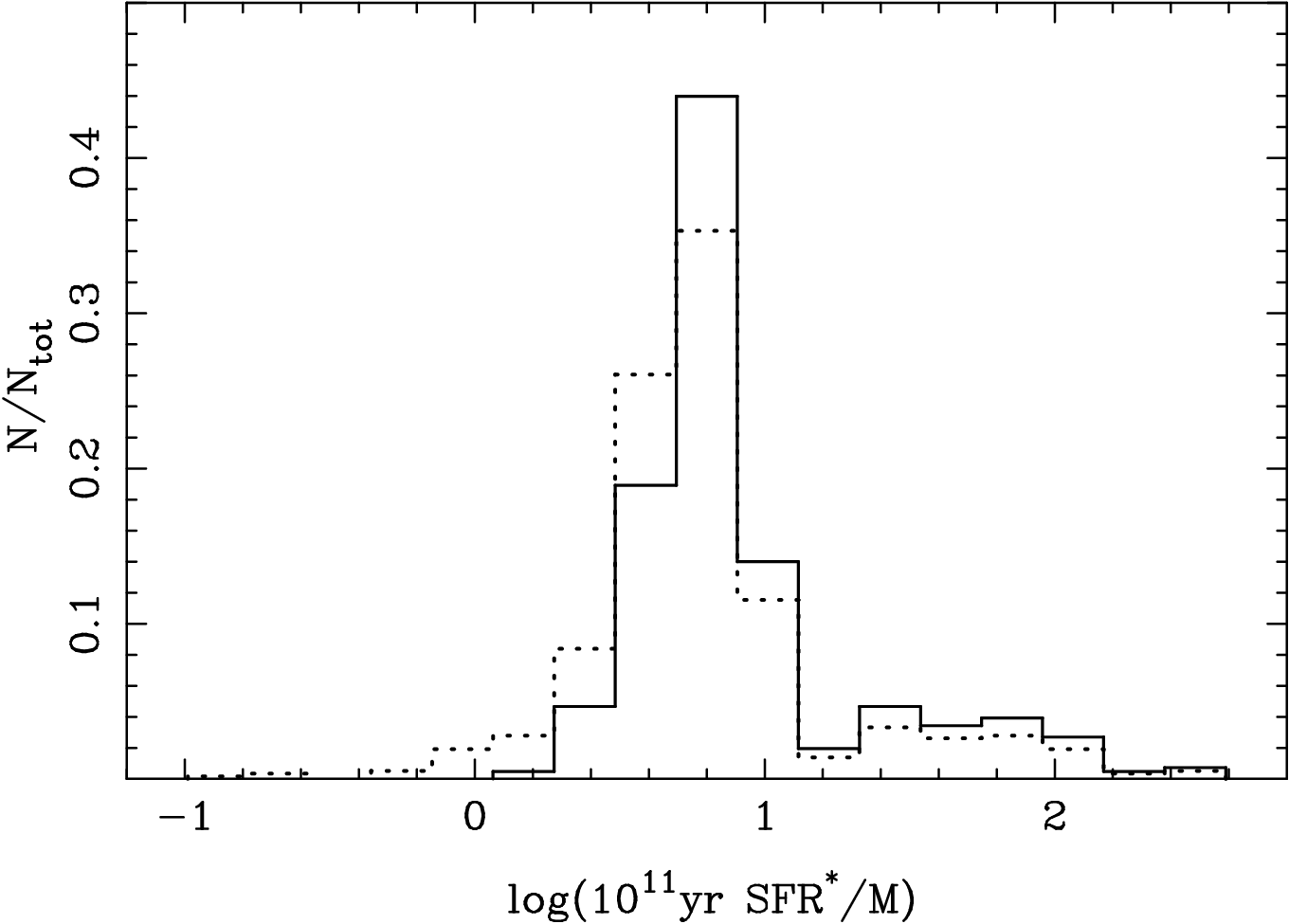}
\caption{Distribution of galaxies per bin of $\epsilon$ for galaxies in
  the catalog at $z=0$. The dotted histograms, representing all galaxies in
  the catalog, are identical in both panels. 
  The solid histogram in the upper panel is for galaxies more
  massive than $5\cdot 10^{10} \ M_{\odot}$, while the solid histogram
  in the lower panel is for objects with epoch of formation less than
  $z=0.9$. All histograms are normalized to the number of objects in
  each distribution.}
\label{fig5}
\end{center}
\end{figure}

We present in figure~\ref{fig5} a statistical quantification of the
star formation efficiency of the galaxy population at $z=0$, where we
have computed the number of objects in a bin of $\epsilon$. The dotted
histogram in both panels represents all galaxies in the catalog. This
distribution peaks around $\epsilon \approx 0.7$, with considerable
tails extending both above and below the peak.  In the upper panel,
the solid histogram shows the corresponding distribution for all
objects more massive than $5\cdot 10^{10} \ M_{\sun}$. The
distribution is now cut off above $\epsilon \approx 1.0$, clearly
demonstrating that it is the lower mass range of objects that have the
highest star formation efficiency at present. The solid histogram in
the lower panel of fig.~\ref{fig5} displays the corresponding
distribution if only recently formed galaxies are included (epoch of
formation less than $z=0.9$). The main difference between the young
population and the total is that the objects with low star formation
efficiency have disappeared. Comparing the two panels, it is apparent
that the low mass objects have the highest star formation efficiency
(upper panel) and the lowest efficiency is generally found in the
oldest objects (lower panel). In the following discussion we will
denote the peak value of a given catalog by $\bar\epsilon$, and refer
to objects with $\epsilon>\bar\epsilon$, as efficiently star-forming
galaxies.  Similarly, galaxies with $\epsilon<\bar\epsilon$ will be
termed inefficient.

\section{Cosmological evolution}
\label{evol}

\begin{figure}
\begin{center}
\includegraphics[height=8.3cm, angle=-90]{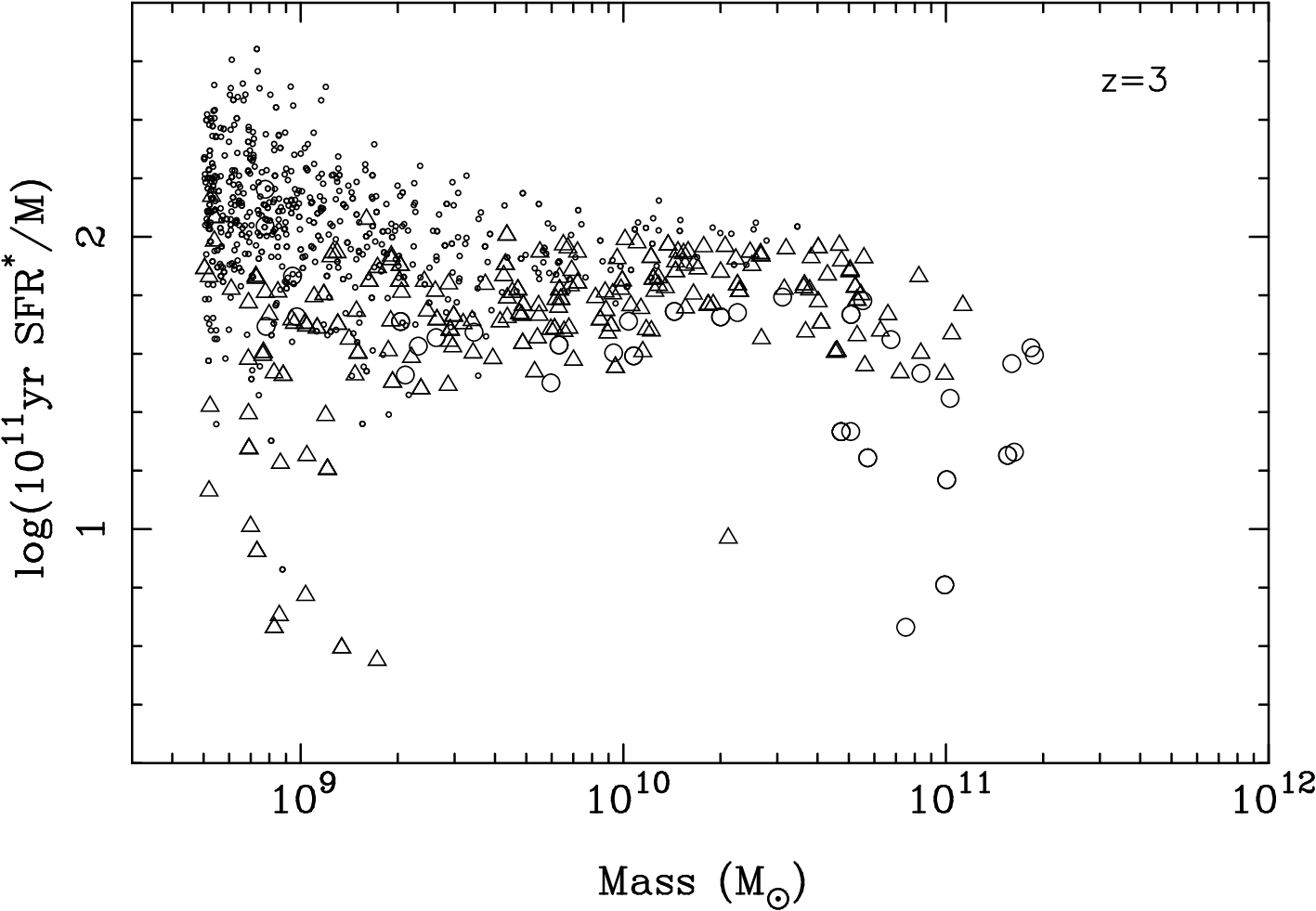}
\includegraphics[height=8.cm, angle=-90]{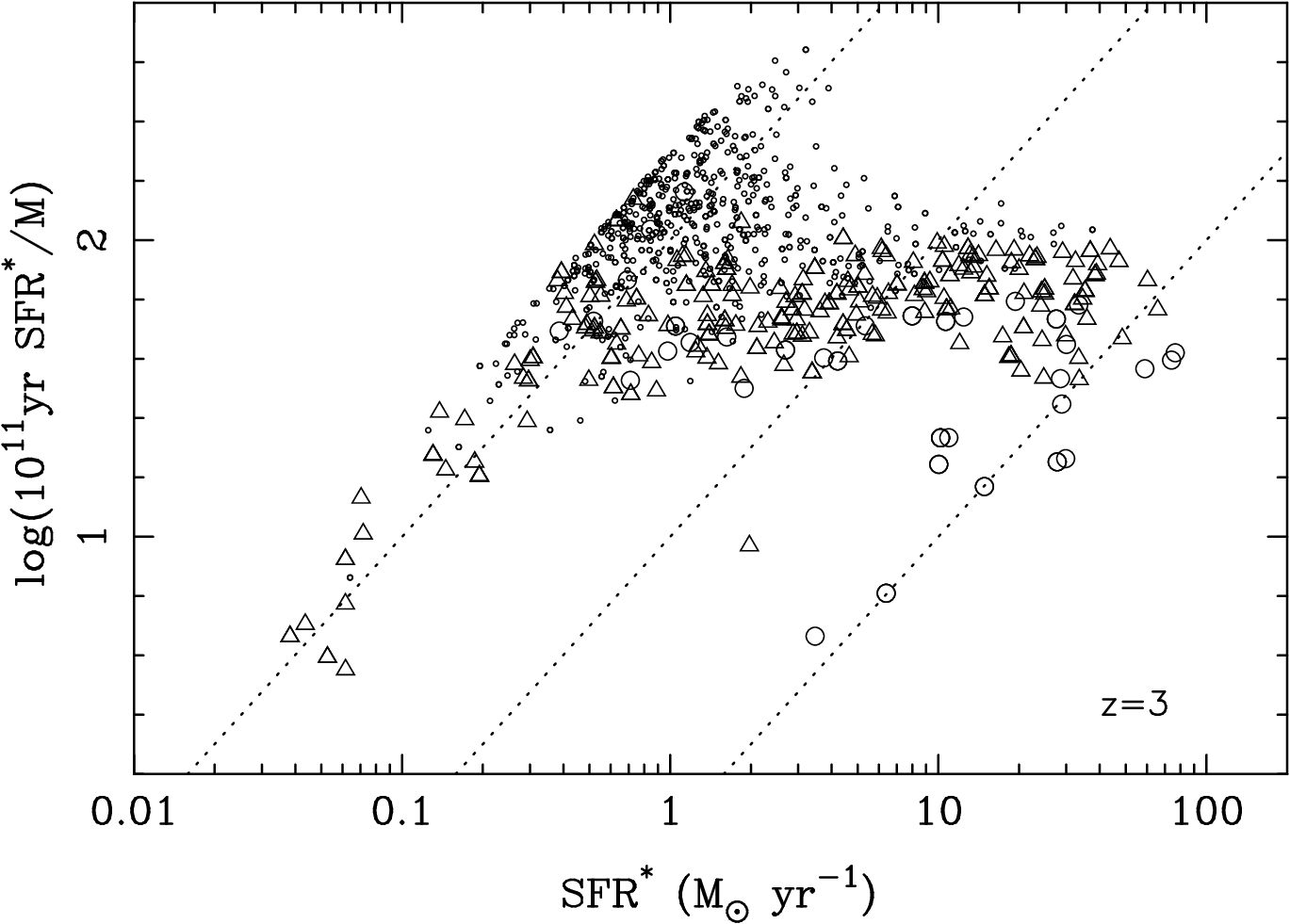}
\caption{Efficiency parameter, $\epsilon$, of galaxies in the 
  catalog at $z=3$ as a function of mass (upper panel), and as a 
  function of $SFR^*$ (lower panel). For clarity, only 1000 objects are 
  plotted in each panel. As in fig.~\ref{fig3}, the dots denote 
  formation epochs lower than $z=3.9$, triangles denote epochs in the 
  range $z=3.9-4.3$, and circles denote formation epochs higher than 
  $4.3$. The dotted lines in the lower panel indicate a constant mass 
  of $10^9, 10^{10}$ and $10^{11}M_{\sun}$ (from left to right).}
\label{fig6}
\end{center}
\end{figure}

The star formation activity of the galaxy population evolves strongly
with redshift as seen in fig.~\ref{fig3}. This evolution is also
manifest in the cosmic evolution of the star formation rate density
(see \cite{Hopkins2001} for a compilation of observational data).
Observations show a strongly decreasing star formation activity
towards lower redshifts for $z\lse 2$, while its evolution towards
high redshift is much less certain.  The actual star formation rate in
a given galaxy not only depends on the history of the galaxy, but also
on merger events, environment, etc.  In this section we explore the
evolution with redshift of the star formation properties discussed in
the previous section. As the cosmic star formation activity decreases
for redshifts below $z=2$, we consider mainly the catalogs of objects
at $z=0$ and $z=3$ for this exploration.

In fig.~\ref{fig6}, we show $\epsilon$ as a function of mass and star
formation rate for the galaxies in the catalog at $z=3$. This should
be compared to fig.~\ref{fig4}, but note the different vertical scale.
The efficiency is essentially constant over most of the mass range and
the greatest dispersion is for objects with mass less than a few times
$10^9M_{\sun}$.  The distribution of $\epsilon$ is much narrower than
at $z=0$, and the mean value is also considerably higher, $\epsilon
\approx 1.8$, an indication of the strong cosmological evolution of
the efficiency of star formation. This trend has also been seen in
observational data \citep{Guzman1997, Brinchmann2000}.  In the lower
panel in fig.~\ref{fig6}, we see that the clear separation of the
galaxies into distinct groups or regions observed at $z=0$, is nearly
absent at $z=3$.  The diagram does show the presence of a
sub-population of low-mass objects with high efficiency but a moderate
$SFR^*$. This population is not clearly separated from the
intermediate mass objects at this redshift, but it continues to be
present at $z=0$ with only slightly lower efficiency and activity.
Comparing with fig.~\ref{fig4}, we note another interesting result,
the appearance at low redshift of the high-mass objects with low
efficiency and low $SFR^*$. Note also that at both redshifts, the
highest mass objects always have the highest $SFR^*$.

\begin{figure}
\begin{center}
\includegraphics[height=8.3cm, angle=-90]{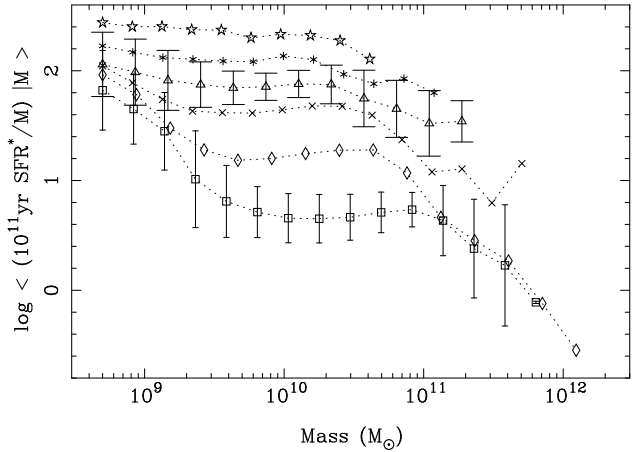}
\caption{Conditional mean of having a specific star formation rate at 
  a given mass, $\langle SFR/M | M\rangle$, as a function of mass at 
  different redshifts: $z=0$ (squares), $z=1$ (diamonds), $z=2$ (crosses), 
  $z=3$ (triangles), $z=4$ (stars), $z=5$ (open stars); for clarity, error 
  bars representing dispersion around the mean are only plotted at $z=0$
  and $z=3$.}
\label{fig7}
\end{center}
\end{figure}

In fig.~\ref{fig7} we show the conditional mean of having a specific
star formation rate at a given mass as a function of mass and its
evolution with redshift. So as not to clutter the diagram too much we
only show the dispersion around the mean for the catalogs at $z=0$ and
$z=3$. In general the dispersion is less at higher redshifts except
for the low-mass objects as already noted in fig.~\ref{fig6}. The
largest difference in $\epsilon$ between low and high-mass galaxies is
seen at low redshift, where only the low-mass and some of the
intermediate mass objects have high specific star formation rate. At high 
redshift the entire galaxy population has a high specific rate. It is 
apparent from the figure that the specific star formation rate of the 
lowest mass objects decreases only slightly from $z=5$ to $z=0$, whereas 
for the intermediate to high-mass objects it decreases by two orders of
magnitudes or more. It is clear that the low-mass objects are not only
the most efficient ones at $z=0$ as already noted, they are efficient
at all redshifts. At higher redshifts the difference between the
efficiency of the different mass ranges diminishes, but the star formation 
efficiency of the low-mass objects is consistent with being constant out 
to a redshift of 5.

\section{Discussion}
\label{discuss}

\begin{figure}
\begin{center}
\includegraphics[height=8.3cm, angle=-90]{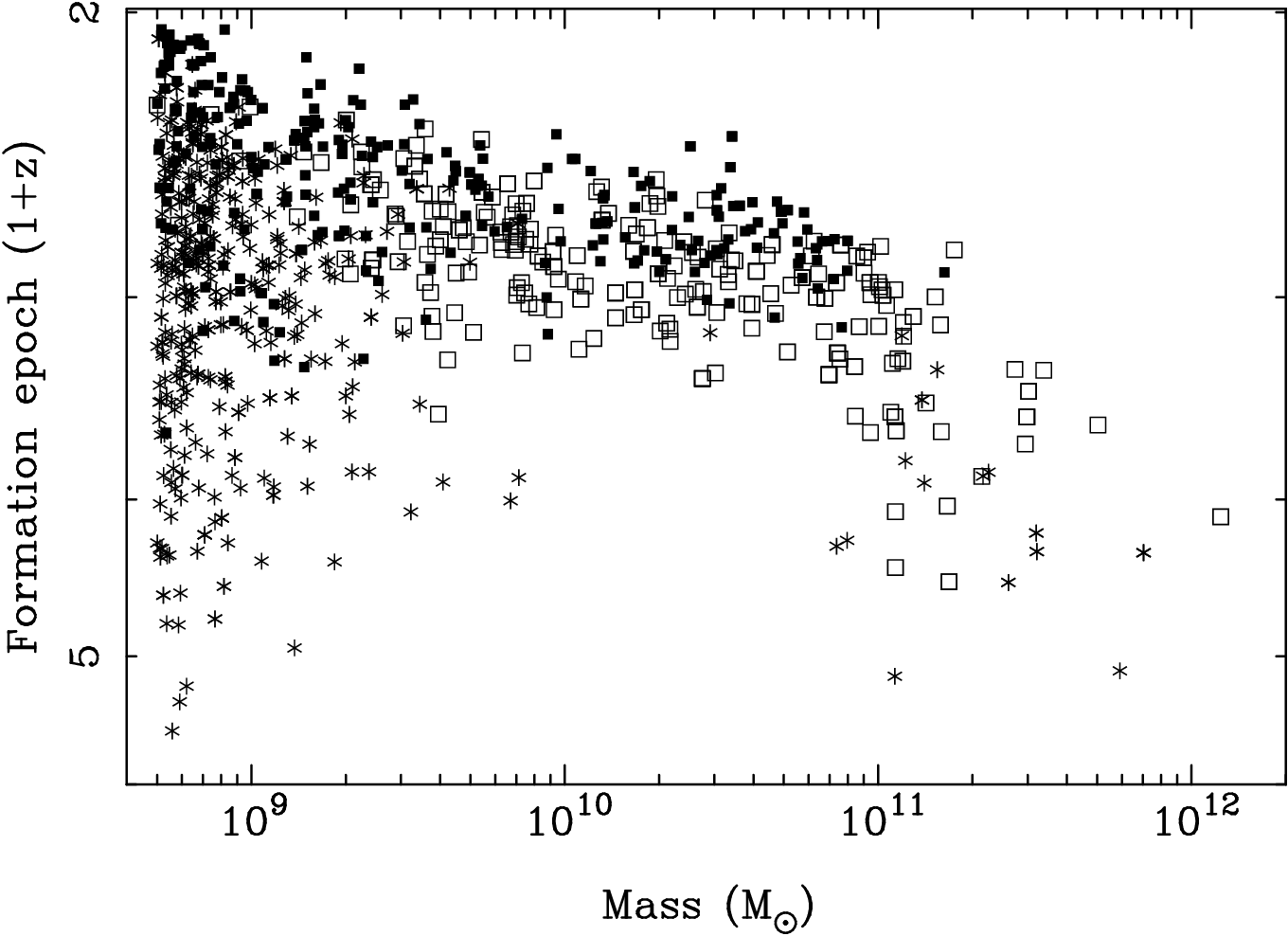}
\includegraphics[height=8.3cm, angle=-90]{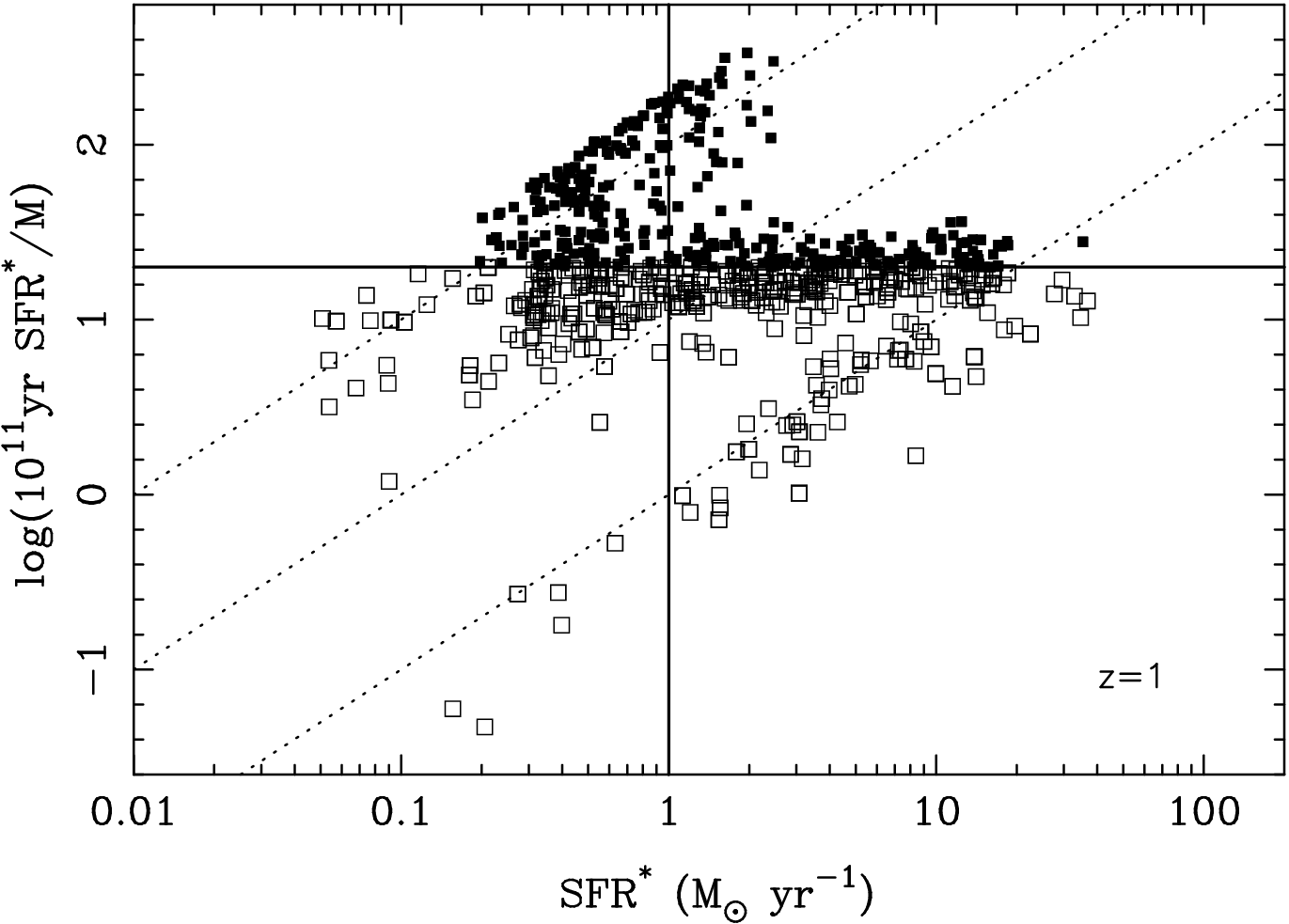}
\caption{Upper panel: Galaxy sub-populations at $z=1$ as a function of 
mass and epoch of formation. Stars denote inactive galaxies ($SFR^*=0$), 
and squares active galaxies ($SFR^*>0$), with empty symbols for inefficient 
and filled ones for efficient galaxies (at this redshift the dividing 
efficiency is $\bar\epsilon=1.3$). Lower panel:
Star formation efficiency, $\epsilon$, of the active population in the catalog 
at $z=1$ as a function of $SFR^*$. For clarity only 1000 randomly selected 
objects are plotted in both panels. As in previous figures, the dotted lines
indicate constant masses of $10^9, 10^{10}$ and $10^{11}M_{\sun}$ (from left 
to right).}
\label{fig8}
\end{center}
\end{figure}

Numerical simulations provide us with catalogs of galaxy-like objects
characterized by different properties such as mass, epoch of formation
and star formation rate. Galaxy sub-populations in the catalogs are
clearly distinct especially at low redshift, e.g.\ the high-mass
population with low star formation rate contrasting with the efficient
low-mass population. The former population shows a decreasing star
formation activity with lower redshifts.  As mentioned in section
\ref{lowz}, a number of objects at low redshift have a zero value of
$SFR^*$ when measured over a time $\tau$. To make the picture of our
simulated galaxy population complete, we show in the upper panel of
fig.~\ref{fig8}, the entire population at $z=1$ as a function of mass
and epoch of formation. In this section we focus on the redshift
$z=1$, as it is close to the peak of the current GRB redshift
distribution and most of the observed host galaxies have redshifts
around unity. The catalog at $z=0$ shows the same trends as is easily
verified by comparing the lower panels in figs.~\ref{fig8} and
\ref{fig4}. In figure~\ref{fig8}, stars refer to the inactive galaxy
population that has no measurable $SFR^*$ at this redshift. Most of
this population is either low-mass galaxies or high-mass galaxies. We
note also that these objects are quite old. Among the active galaxy
population (square symbols), galaxies with a non-zero $SFR^*$ (1239
objects) are separated in the upper panel of fig.~\ref{fig8} into an
efficient population, including all galaxies with
$\epsilon>\bar\epsilon=1.3$ (filled squares), and an inefficient
population with $\epsilon<\bar\epsilon=1.3$ (open squares). The most
efficient galaxies are low-mass and late-formed galaxies as already
pointed out in the previous sections. The lower panel of
fig.~\ref{fig8} focuses on the active population in a $\epsilon$ vs.\ 
$SFR^*$ diagram. This plot (as the one at $z=0$ in fig.~\ref{fig4})
reveals again that high star formation rate galaxies are not
necessarily galaxies with the highest specific $SFR^*$. This kind of
representation emphasizes the distinction between the star formation
activity ($SFR^*$, along the horizontal axis) and the star formation
efficiency ($\epsilon$, along the vertical axis), as two different
measures of the process of converting baryons into stars.

\begin{figure}
\begin{center}
\includegraphics[height=8.3cm, angle=-90]{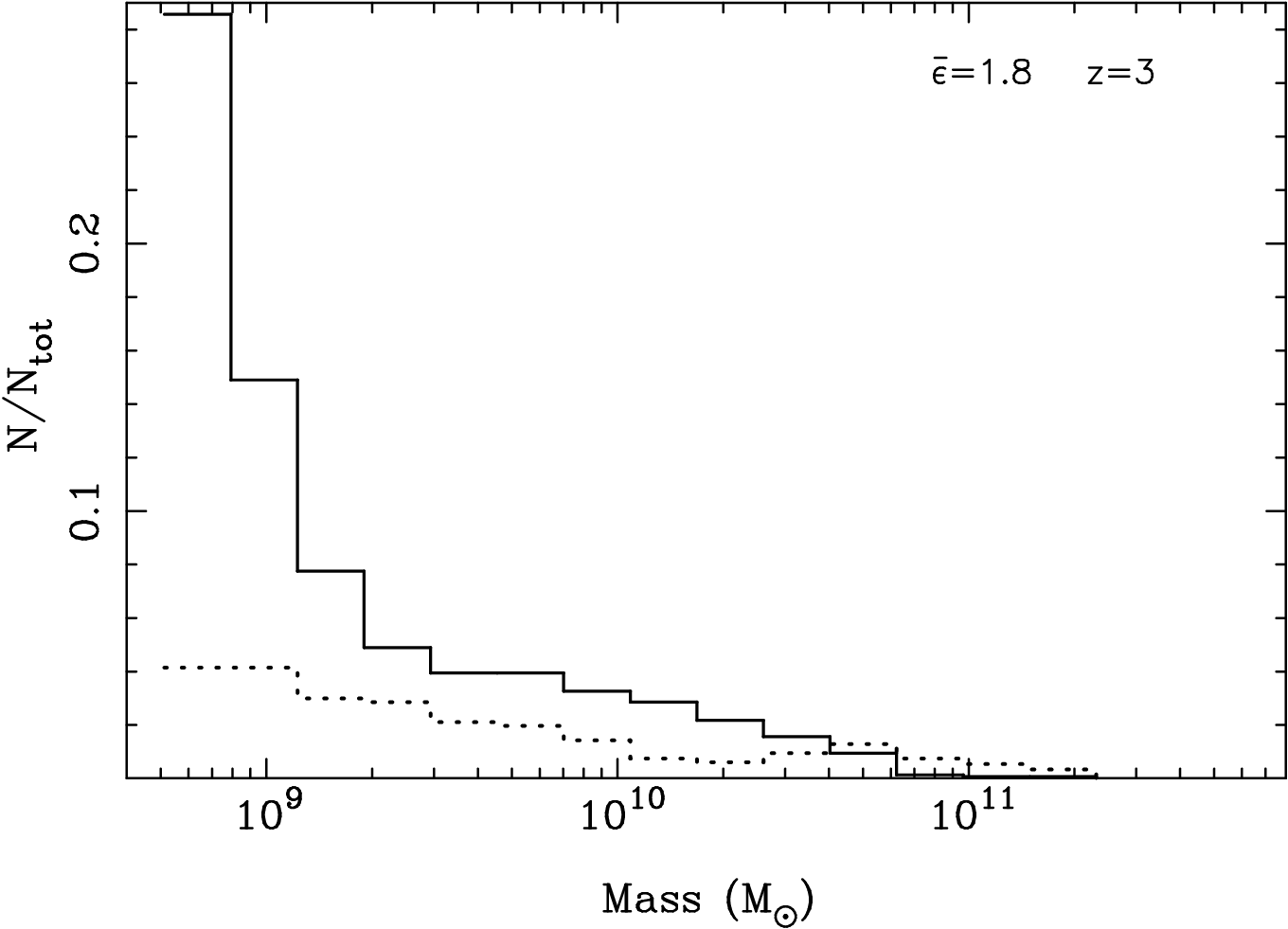}
\includegraphics[height=8.3cm, angle=-90]{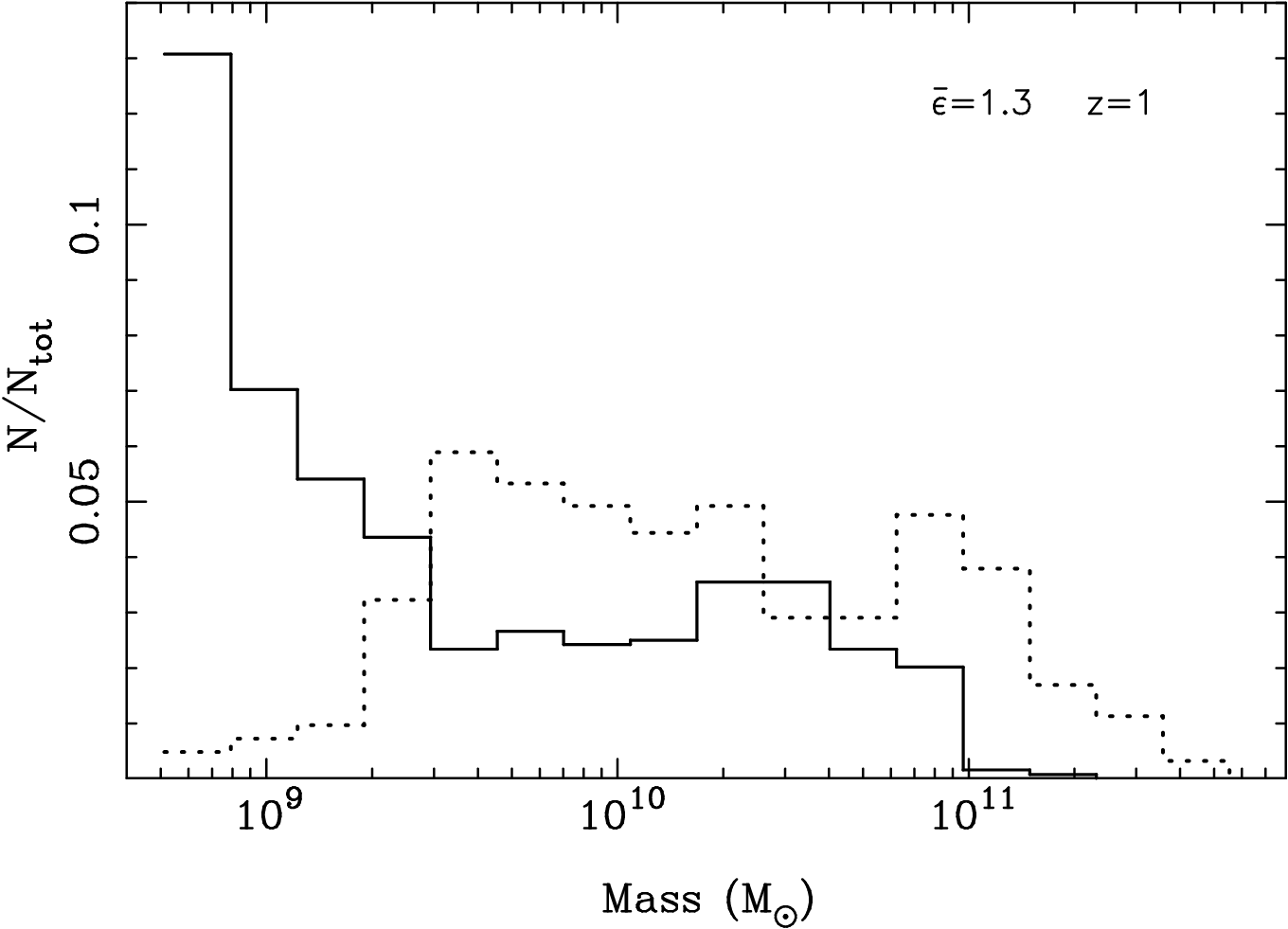}
\caption{Number of galaxies per bin of mass at redshift of $z=3$
  (upper panel) and $z=1$ (lower panel). In each panel the solid
  histogram shows the distribution of the efficient population (with
  $\epsilon>\bar\epsilon=1.8$ at $z=3$ and $\epsilon>\bar\epsilon=1.3$ 
  at $z=1$) and the dotted histogram shows the non-efficient population 
  (with $\epsilon<1.8$ at $z=3$ and $\epsilon<1.3$ at $z=1$). Both 
  histograms are normalized on the whole population.}
\label{fig9}
\end{center}
\end{figure}

The efficient and inefficient populations are statistically
characterized at high and low redshifts in fig.~\ref{fig9} where we
show their distribution per bin of mass. Note that both distributions
at each redshift are normalized on the whole population, so adopting
lower or higher values of $\bar\epsilon$ would alter the shape of both
the efficient and inefficient populations.  At $z=3$, most of the
objects with star formation efficiency above $\bar\epsilon=1.8$ (solid
histogram) are low-mass objects ($M<$ a few times $10^9\, M_{\sun}$).
High-mass objects are not very numerous at this redshift (cf.\ 
fig.~\ref{fig6}), and most of them are inefficient. At $z=1$ most of
the efficient objects are again predominantly of low-mass, although a
considerable fraction of intermediate mass objects (approximately
$10^{10}-10^{11}M_{\sun}$) also have high efficiency.  The
non-efficient objects at $z=1$, are dominated by the mass range of few
times $10^9\, M_{\sun}$ to almost $10^{12}\, M_{\sun}$. At the lower
redshift, essentially all of the low-mass objects are efficient in
their star formation.  The same can be said about the corresponding
distribution at $z=0$, almost all low-mass objects are efficient. To
summarize, we note that efficient galaxies exist over almost the
entire mass range at high redshift (see also fig.~\ref{fig7}). At low
redshift the efficiency becomes strongly mass dependent, as is also
clear from the lower panel in fig.~\ref{fig4}. Most importantly, the
efficient population is dominated by a large number of low-mass
objects, independent of the redshift. Considering the lower panels of
figs.~\ref{fig6} and \ref{fig8}, we clearly see that the non-efficient
objects have a larger spread in their $SFR^*$ than the efficient
galaxies.  The $SFR^*$ of the latter sub-population are clustered at
moderate values.

The star formation efficiency, $\epsilon$, of a given galaxy is
expected to be closely related to its color. Indeed, the instantaneous
$SFR$ is dominated by the youngest stellar populations and thus is a
measure of the luminosity in the optical/UV bands. The galaxy mass is
on the other hand a better measure of the luminosity in the infrared
band, assuming some mass-to-light ratio, since the mass is the
integration of all stellar populations produced over the lifetime of
the galaxy. Low-mass galaxies of the simulated catalogs have the
highest specific $SFR^*$ and are thus expected to be bluer than
massive galaxies; observations of compact galaxies show bluer colors
and lower mass than high-mass disk-like starburst galaxies
\citep{Guzman1997}.

The first observational results on GRB host galaxies provide important
clues to their nature. \cite{LeFloch2003} showed that the hosts are
faint in the $K$ band, typical luminosities being about $0.1L_*$.
\cite{Berger2003} have shown that the hosts are bluer than other
sub-mm galaxies. \cite{Chary2002} considered 11 GRB host galaxies and
derived extinction corrected $SFR$ for seven of these hosts as well as
mass, starburst age and internal extinction.  Six of them have
redshifts between 0.7 and 1.6, one is at a high redshift of $z=3.4$.
They derive UV and $\beta$-slope $SFRs$ in the range 0.2-70
$M_{\sun}$/yr for the hosts with redshifts around unity and more than
300 $M_{\sun}$/yr for the host with highest redshift. Using population
synthesis spectral energy distributions they computed the mass of host
galaxies and found them to be of rather low-mass, between $10^8$ and
$4 \cdot 10^9 M_{\sun}$ for the hosts with redshifts around unity. The
host galaxy at $z=3.4$ is a higher mass object, $3\cdot 10^{10}
M_{\sun}$. These mass values are also consistent with other estimates
\citep{Sokolov2001}.  As a result, \cite{Chary2002} estimate the
specific star formation rate of these hosts, approximately in the
range $\epsilon=2.5-3.9$ in our units (see their fig.~2). Even with
some uncertainties, these values are consistent with values inferred
for star-bursting galaxies (see upper panel of fig.~16 in
\cite{PerGon2003} based on observations by \cite{Calzetti97}).  Future
observations will reveal more details of the host galaxies and
especially the connection these have with the observed population of
blue galaxies, which could be irregulars, star-bursting, compact or
HII.

Based on these host galaxy observations and on the fact that we can
identify galaxy sub-populations in the simulations, we define our
simulated candidate GRB host galaxies as efficiently star-forming
objects, with high specific star formation rate (filled squares in
fig.~\ref{fig8}) A galaxy is considered a host if its efficiency is
higher than $\bar\epsilon$, taken as the peak value of a given
catalog. This value is rather arbitrary and is likely to be a lower
limit, according to the observational estimate of the specific star
formation rate in \cite{Christensen2004}. The majority of these
efficient galaxies in the simulations are low-mass objects, but a
modest peak in the mass distribution around a few times $10^{10}
M_{\sun}$ is also seen in the lower panel of fig.~\ref{fig9}.  Such
galaxies are expected to be around 10 times less luminous than $L_*$
galaxies, consistent with \cite{LeFloch2003}. Moreover, the simulated
host galaxy candidates are moderately active, with $SFR^*$ spanning
two orders of magnitude around unity. Their specific star formation
rates spans about one order of magnitude (fig.~\ref{fig8}).  Most of
these galaxies are also low-mass objects, clustering around the line
representing $10^9M_{\sun}$. This picture seems consistent with
observations: Although the host galaxies in \cite{Chary2002} show much
higher specific star formation rates than our candidates, it varies by
a factor of 20 and the host masses are generally around $10^9 \ 
M_{\sun}$. It is also interesting to note that one of their hosts with
$z=3.4$, has a similar $\epsilon$ as the hosts around $z=1$, but with
a much higher star formation rate and a galaxy mass that is two orders
of magnitude higher. The simulations show a similar result, a $10^9 \ 
M_{\sun}$ galaxy has a specific star formation rate in the range
$\epsilon\approx 1.3-2.6$ at $z=1$ (see fig.~\ref{fig8}), which
brackets the values obtained for a $10^{10} \ M_{\sun}$ galaxy at
$z=3$ (fig.~\ref{fig6}). Recall also, that the $SFR^*$ increases at
higher redshift for a given mass.

\begin{figure}
\begin{center}
\includegraphics[height=8.3cm, angle=-90]{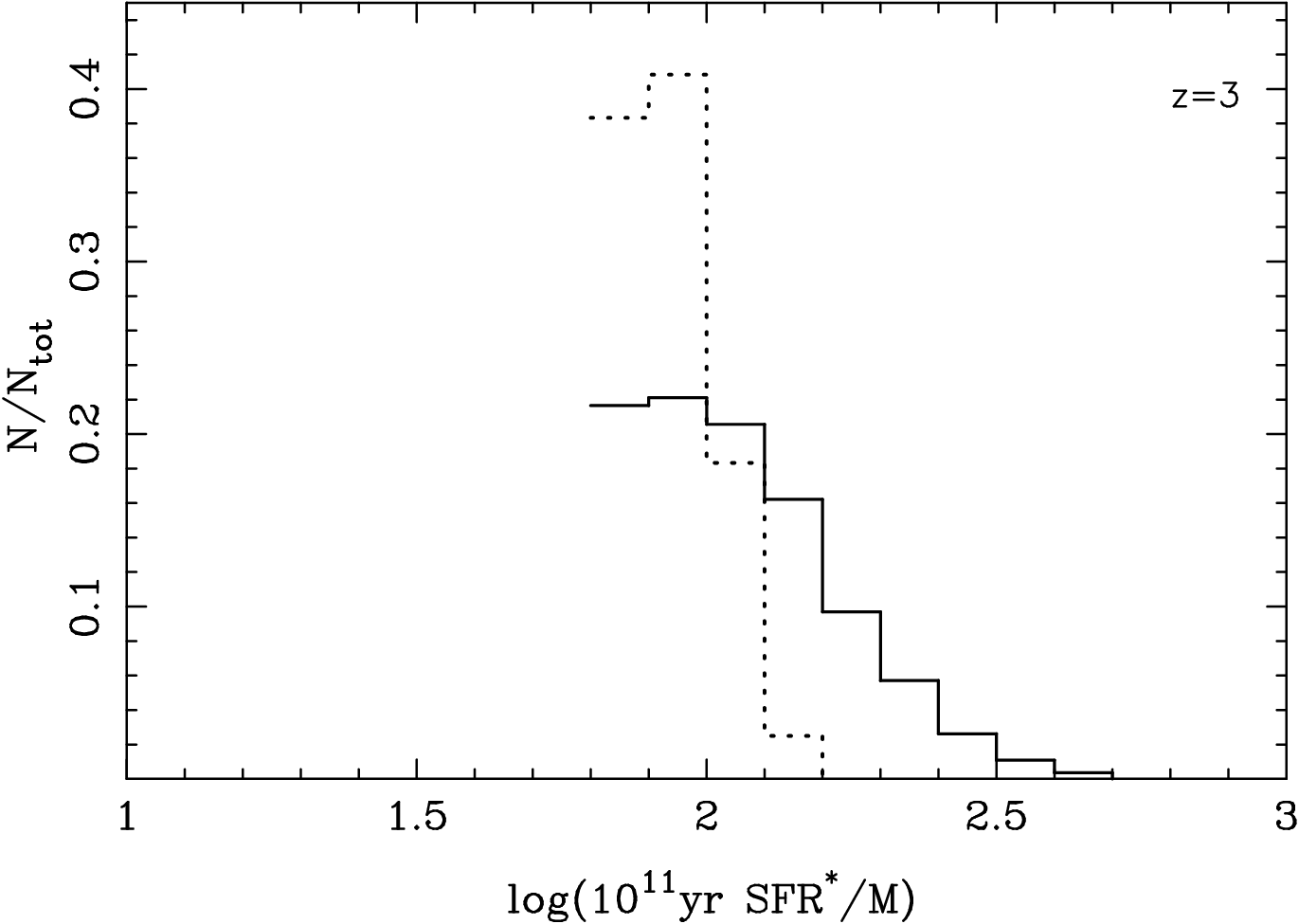}
\includegraphics[height=8.3cm, angle=-90]{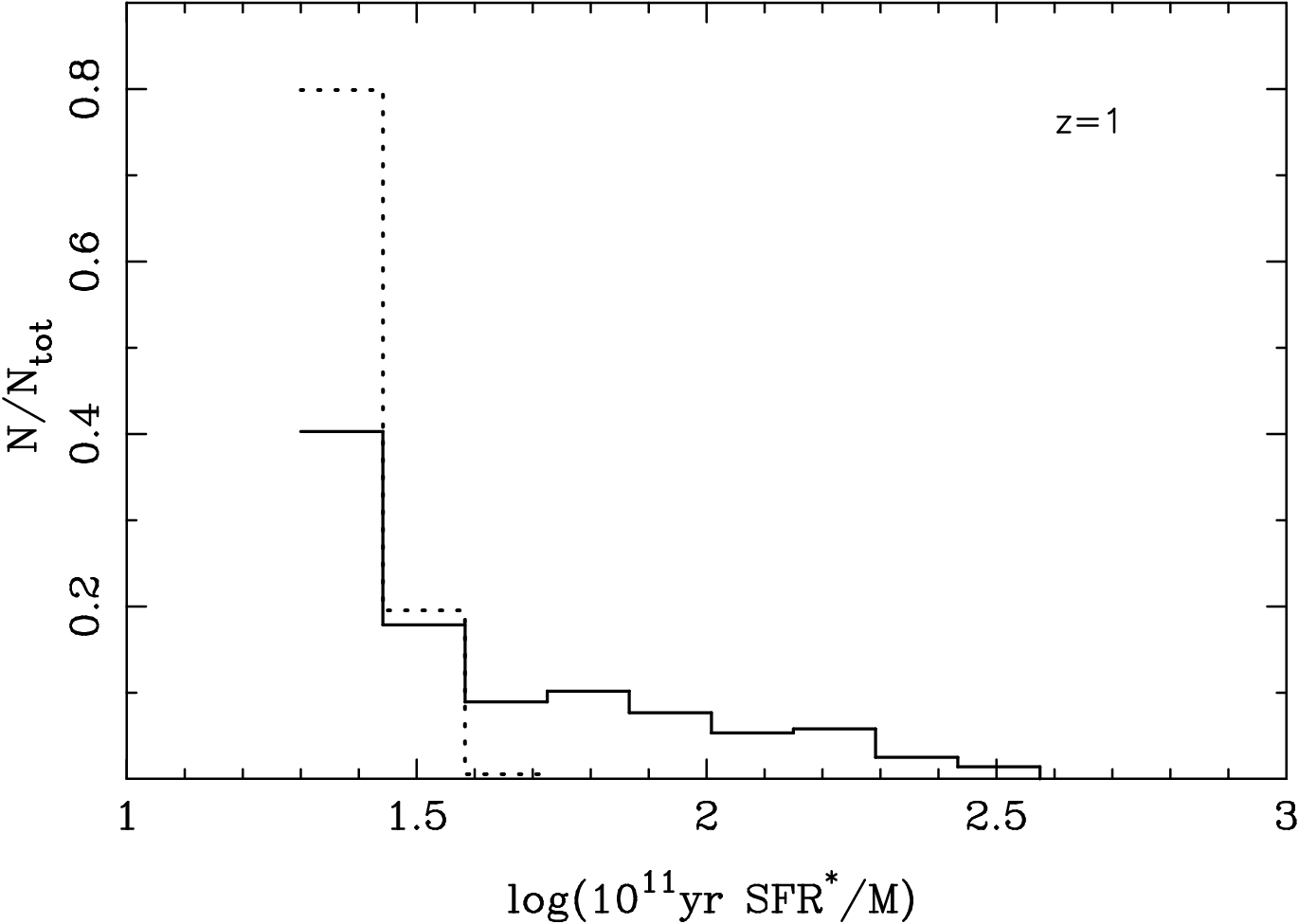}
\caption{Upper panel: Number of host galaxies ($\epsilon>\bar\epsilon=1.8$)
  per bin of specific star formation rate (solid histogram) compared
  to the distribution of host galaxies with $M >10^{10} M_{\sun}$
  (dotted histogram) at $z=3$. Lower panel: same but at $z=1$ 
  with $\epsilon>\bar\epsilon=1.3$.
  Histograms are normalized to the number of objects in each
  distribution.}
\label{fig10}
\end{center}
\end{figure}

The location of the candidate host galaxies in the $\epsilon-SFR^*$
parameter space, appears to indicate that most of them would have high
$\epsilon$ and modest to high $SFR^*$. This is also what the
observations of \cite{Chary2002} show. Moreover, a number of hosts are
classified as compact galaxies, and such galaxies appear to have $SFR$
between 0.1 and 14 $M_{\sun}$/yr as in the sample of
\cite{Guzman1997}.  Putting this together suggests that identifying
host galaxies of GRBs as efficient galaxies provides us with first
interpretations of the observations.  Our results also point to a
verifiable prediction regarding the observationally determined
distribution of the specific star formation rates of GRB hosts.
Observations of host galaxies could, however, be biased towards the
most luminous objects and therefore the intermediate to high-mass
range, with the more numerous and efficient, but much less massive,
objects underrepresented.  The solid histograms in fig.~\ref{fig10}
show the efficiency distribution of the candidate host galaxies (i.e.\ 
objects with $\epsilon>\bar\epsilon$) at high and low redshifts.
Considering all efficient objects results in broad distributions
(solid histograms).  If we now restrict the hosts to be the most
massive of the efficient objects ($M>10^{10}M_{\sun}$), we obtain
strikingly different distributions.  They are much narrower (dotted
histograms) at both redshifts with the tail towards higher efficiency
missing. First versions of observationally determined distributions
similar to the dotted ones in fig.~\ref{fig10} already exist
\citep{Christensen2002}.  The difference between the histograms,
illustrate how our perception of the hosts will change if we are able
to detect these low-mass galaxies.
\\

The conclusion that host galaxies are, in majority, low-mass,
late-formed galaxies with moderate star formation rates, derives from
our hypothesis that candidate host galaxies are the most efficient
ones. We adopted this hypothesis as observations indicate that the
hosts have high specific star formation rates and moderate $SFR$
\citep{Christensen2004}. Our conclusions would be different if the
hypothesis is that candidate hosts are galaxies with the highest star
formation rates, as may be expected if the GRB formation rate follows
the global $SFR$. In that case, the majority of the host galaxies
would be high-mass, inefficient, early-formed galaxies (see
Fig.~\ref{fig8}).  It may be argued that for our adopted hypothesis to
hold, and if GRBs trace the global $SFR$, low-mass galaxies should
contribute significantly to the global star formation rate. However,
inspection of Fig.~\ref{fig8}, immediately shows that the matter is
more complicated. If we select a sample of objects according to a
particular criteria, we find that at $z=1$, 50$\%$ of the total star
formation rate is in galaxies with a $SFR^*< 9\; M_{\sun}$/yr; or in
galaxies with $\epsilon>1.3$; or in galaxies with a mass less than
$5\cdot 10^{10}M_{\sun}$.
Similarly at $z=3$, these thresholds become $SFR^*=17$ $M_{\sun}$/yr;
or $\epsilon=1.9$; or $M=2\cdot 10^{10}M_{\sun}$.
Of course we have to keep in mind that these numerical values 
depend on the physics included in our simulations.
  
It is clear from Fig.~\ref{fig8} that if a higher specific star
formation threshold is used to select candidate host galaxies, fewer
hosts will have high $SFR^*$s. This would suggest that the GRB
formation rate is not a tracer of the total star formation rate. The
two rates follow each other (i) if the stellar mass function is
independent of time and thus of the galaxy properties (such as mass,
age, metal content, etc), and (ii) if the probability that a high-mass
star gives birth to a GRB event is independent of time. Although a
number of observational studies show that the stellar mass function
depends on the galaxy type \citep{Contini95, Kroupa2001, Larson2003},
most workers assume a universal initial mass function independent of
time. Moreover, the underlying physics involved in the gamma-ray burst
phenomena implies, among other things, that low-metalicity stars are
favored. Such low-metalicity environments should mainly exist in
young or starbursting or interacting galaxies rather than in evolved
systems even if these may be forming high-mass stars. Similar caveats
have been discussed in other contexts by \cite{Ram-Ruz2002} and
\cite{Choudhury2002}.  Observing host galaxies with properties such as
a high specific star formation rate, could reinforce the fact that the
probability of forming a GRB event from a high-mass star does depend
on time and thus on the galaxy.

\section{Conclusions}
\label{conclude}

We have used numerical simulations of large scale structure formation
to identify galaxy like objects and to follow their evolution. We
concentrated on the mass, the epoch of formation and the instantaneous
and specific star formation rates of the objects. Our most important
results are as follows:

The whole galaxy population includes a significant number of low-mass
objects with $M<10^{10}M_{\sun}$. There is a strong cosmological
evolution of the properties of the galaxy population between $z=3$ and
$z=0$. We emphasize that the star formation rate of a galaxy should be
considered a measure of its star formation {\em activity}, while the
specific star formation rate is a measure of its star formation {\em
  efficiency}. These properties are nicely illustrated in a diagram of
$\epsilon$ vs.\ $SFR^*$ (fig.~\ref{fig8}), where we find a number of
clearly distinct sub-populations of galaxies.

The low-mass population is further divided into sub-populations, some
of which have a high specific star formation rate. Based on these
findings and on the first observational results of GRB hosts, we
identify candidate hosts in the simulations as galaxies with high
specific $SFR^*$, rather than objects with high star formation rate.
The majority of these candidate host galaxies appear to be of low
mass, with recent formation epochs and a moderate $SFR^*$ of a few
$M_{\sun}$/year or lower. The properties of the candidate hosts so
identified, are consistent with the trends observed in the GRB host
galaxies. An implication of our conclusions, given the applicability
of the physical input in the simulations used here, is that GRBs may
not trace the cosmic star formation rate.

Most of the observed GRB hosts to date have a redshift around unity.
We show that the efficiency of the low-mass galaxies does not vary
much with redshift.  For high redshift, more and more galaxies are
found to be efficient independent of mass, while the efficiency of
intermediate to high-mass galaxies decreases strongly with decreasing
redshift.

As the low-mass objects are very faint, the observations of GRB host
galaxies will initially be observationally biased towards the
intermediate and high-mass objects. These will be easier to detect due
to their higher luminosity. They will typically have an $SFR$ in the
range $1-40 M_{\sun}$/year at a redshift of unity. It may take the
next generation space telescopes to reveal the properties of the
low-mass sub-population of host galaxies. These are expected to
outnumber the currently observable hosts by a considerable factor.

Furthermore, the low-mass galaxy population is known to be important
for the understanding of galaxy formation. GRB host galaxies may thus
become useful tracers of this population or in fact the faint end of
the galaxy luminosity function that may be difficult to detect
otherwise. Thus GRB selected host galaxies may become an important
link in the study of structure formation and evolution. Although GRB
selected galaxies introduce its own 'selection effect', they are free
from others affecting surveys of various kinds, such as the magnitude
limited observations of optical surveys and the sensitivity limitation
of the sub-mm observations.

\section*{Acknowledgments}
We thank the  anonymous referee for constructive comments that helped
improve the paper. This work was supported by a Special Grant from the 
Icelandic Research Council. The numerical simulations used in this 
paper were performed on NEC-SX5 at the Institut du D\'eveloppement 
et des Ressources en Informatique Scientifique (France).

\bibliography{paper}

\end{document}